\documentclass[twocolumn,trackchanges]{aastex62}
\usepackage{natbib}
\usepackage{amsmath}
\DeclareUnicodeCharacter{2061}{}
\received{January 1, 2022}
\revised{January 7, 2022}
\accepted{\today}
\submitjournal{ApJS}

\shorttitle{BHB stars}
\shortauthors{Ju et al.}
\usepackage{amsmath}
\begin{document}

\title{The Blue Horizontal-Branch Stars From 
 the LAMOST Survey: Atmospheric Parameters}

\correspondingauthor{Wenyuan Cui}
\email{wenyuancui@126.com, cuiwenyuan@hebtu.edu.cn}

\author{Jie Ju}
\affil{Department of Physics, Hebei Normal University, Shijiazhuang 050024, China}
\affil{Guo Shoujing Institute for Astronomy, Hebei Normal University, Shijiazhuang 050024, China}
\affil{School of Sciences, Hebei University of Science and Technology, Shijiazhuang 050018, China}

\author{Bo Zhang}
\affiliation{Key Laboratory of Space Astronomy and Technology, National Astronomical Observatories, Chinese Academy of Sciences, Beijing 100101, China}

\author[0000-0003-1359-9908]{Wenyuan Cui}
\affil{Department of Physics, Hebei Normal University, Shijiazhuang 050024, China}
\affil{Guo Shoujing Institute for Astronomy, Hebei Normal University, Shijiazhuang 050024, China}

\author{ZhenYan Huo}
\affil{Department of Physics, Hebei Normal University, Shijiazhuang 050024, China}
\affil{Guo Shoujing Institute for Astronomy, Hebei Normal University, Shijiazhuang 050024, China}

\author{Chao liu}
\affiliation{Key Laboratory of Space Astronomy and Technology, National Astronomical Observatories, Chinese Academy of Sciences, Beijing 100101, China}
\affiliation{Institute for Frontiers in Astronomy and Astrophysics, Beijing Normal University, Beijing 102206, China}
\affiliation{University of Chinese Academy of Sciences, Beijing 1014086, China}

\author{Yang Huang}
\affiliation{School of Astronomy and Space Science, University of Chinese Academy of Sciences, Beĳing 100049,  China}
\affiliation{Key Laboratory of Optical Astronomy, National Astronomical Observatories, Chinese Academy of Sciences, Beijing 100101, People’s Republic of China}

\author{JianRong Shi}
\affiliation{Key Laboratory of Optical Astronomy, National Astronomical Observatories, Chinese Academy of Sciences, Beijing 100101, People’s Republic of China}
\affiliation{School of Astronomy and Space Science, University of Chinese Academy of Sciences, Beĳing 100049,  China}

%
%



\begin{abstract}
Blue horizontal-branch (BHB) stars are crucial for studying the structure of the Galactic halo. Accurate atmospheric parameters of BHB stars are essential for investigating the formation and evolution of the Galaxy. In this work, a data-driven technique named stellar label machine (SLAM) is used to estimate the atmospheric parameters of Large Sky Area Multi-Object Fiber Spectroscopic Telescope low-resolution spectra (LAMOST-LRS) for BHB stars with a set of A-type theoretical spectra as the training dataset. We add color indexes ($(BP-G), (G-RP), (BP-RP), (J-H)$) during the training process to constrain the stellar temperature further. Finally, we derive the atmospheric parameters ($T_\mathrm{eff}$, log\, $g$, [Fe/H]) for 5,355 BHB stars. Compared to existing literature results, our results are more robust, after taking the color index into account, the resulted precisoin of $T_\mathrm{eff}$, log\, $g$ is significantly improved, especially for the spectrum with low signal-to-noise ratio (S/N). Based on the duplicate observations with a S/N difference $< 20\%$, the random errors are around 30\,K, 0.1~dex, and 0.12~dex for $T_\mathrm{eff}$, log\,$g$, [Fe/H], respectively. The stellar labels provided by SLAM are also compared to those from the high-resolution spectra in literature. The standard deviation between the predicted star labels and the published values from the high-resolution spectra is adopted as \sout{to} the statistical uncertainty of our results. They are  $\sigma$($T_\mathrm{eff}$) = 76\,K, $\sigma$(log\,$g$) = 0.04~dex, and $\sigma$([Fe/H]) = 0.09~dex, respectively.

\end{abstract}



\section{Introduction} 

BHB stars are characterized by slow rotation and low mass. They are classified into three types based on their temperature: the A-type horizontal branch ($T_{\rm eff}$$ <$ 12,000\,K), the B-type horizontal branch (12,000\,K  $<$$T_{\rm eff}< $20,000\,K), and the extreme or extended horizontal branch \citep[$T_{\rm eff} >$ 20,000\,K;][]{2009Ap&SS.320..261C}.

Most BHB stars are found in the galactic halo. BHB stars are known for their luminosity and nearly constant absolute magnitude, making them valuable as `standard candle' stars commonly utilized in the study of the Galactic halo \citep[e.g.,][]{1974ApJS...28..157G,1992AJ....103..267B,2000AJ....120.1579Y}. BHB stars, known for being metal-poor population II stars, are ideal for studying the early evolution of the Galaxy \citep{2009ApJ...700.1282Y,2010ApJ...712..516N,2019MNRAS.490.5757S} and are crucial for estimating the mass of halo \citep{1991ApJ...380..104N,2008ApJ...684.1143X}. The stellar atmospheric parameters of BHB stars, such as effective temperature ($T_{\rm eff}$), surface gravity (log\,$g$), and metallicity ([Fe/H]), play a crucial role in determining the distance and age of these stars. Predicting the atmospheric parameters is an essential step for further studies.
   
The effective temperatures of BHB stars are higher than $T_{\rm eff}>$7000\,K, they have strong Balmer lines, while weaker spectral features compared to the cooler stars \citep{2009ssc..book.....G}. It is quite challenging to predict atmospheric parameters with so few spectral features. Therefore, obtaining their atmospheric parameters only through the spectrum is insufficient.  Meanwhile, the atmospheric diffusion affects the metallicity and surface gravity measurements of hot BHB stars ($T_{\rm eff}$$ >$ 11,000\,K )\citep{behr2003chemical,liu2023origins,culpan2024probing}.

\citet{1999AJ....117.2308W} conducted a study on BHB and A-type stars, and obtained $T_{\rm eff}$ (from 6000\,K to 10,000\,K), $\log{g}$, and metallicity using {\it UBV} colors, Balmer line profiles, as well as equivalent widths of \ion{Ca}{2}\,K. 
 \citet{2000A&A...364..102K} estimated the stellar parameters of field BHB stars using moderately high-resolution spectra and IUE ultraviolet low-resolution spectra, as well as available broad-band photometry.
\citet{2003ApJS..149...67B}  studied red and blue field horizontal branch Stars using high-resolution spectra with both the photometric and spectroscopic techniques. Moreover, \citet{2008AJ....136.2022L} estimated the atmospheric parameters of AFGK-type stars with the SEGUE Stellar Parameter Pipeline (SSPP) using low-resolution spectroscopy and {\it ugriz} photometry from low-resolution spectra from the Sloan Digital Sky Survey (SDSS-I) \citep{2000AJ....120.1579Y} and its Galactic extension (SDSS-II/SEGUE). However, these studies only included a limited sample size.  Meanwhile, most studies have focused on the atmospheric parameters of A-type stars. 

With the rise of machine learning algorithms, a data-driven approach has become more efficient in estimating stellar atmospheric parameters for large amounts of spectral data. One such model is the Cannon, developed by \citet{2015ApJ...808...16N}, which estimated the stellar parameter for a large set of 55,000 stars from the APOGEE DR10. Similarly, \citet{2021ApJS..253...22X} constructed a neural network to derive the absolute magnitude for 16,002 OB stars. \citet{2020ApJS..246....9Z} used the SLAM Stellar Label Machine, a data-driven learning module trained by APOGEE stellar parameters, to estimate the stellar label for K-giants. \citet{2021ApJS..253...45L} used the SLAM to obtain the stellar labels of M dwarfs using the spectra from the Large Sky Area Multi-Object Fiber
Spectroscopic Telescope medium-resolution survey (LAMOST-MRS). \citet{2021ApJS..257...54G} adopted SLAM to calculate the stellar labels for early-type stars from LAMOST
 spectra. Lastly, \citet{2022A&A...662A..66X} estimated 330,000 OBA stars from LAMOST DR6 using the HOTPAYNE method.

Large surveys such as SDSS/APOGEE and LAMOST have gathered a massive amount of spectra, leading to the identification of a large sample of BHB stars \citep{2011ApJ...738...79X,2019MNRAS.490.5757S,2021ApJ...912...32V,2021A&A...654A.107C}. However, the stellar parameters have not been provided by these catalogs. 
 Therefore, estimating the stellar parameters of a large sample of BHB stars with low-resolution spectra is crucial. 
This study uses SLAM to establish a catalog of BHB stars with reliable stellar atmospheric parameters. These atmospheric parameters will be very useful in the study of Galaxy evolution.

In this article, we describe the sample of LAMOST-LRS spectra in Section~2, and a concise introduction to how SLAM operates is provided in Section~3. Section~4 is the results and validation, and finally, the conclusion is given in Setion~5.

\section{DATA}

LAMOST is a unique 4m  quasi-meridian reflecting Schmidt telescope, the LAMOST low-resolution spectra cover a wavelength range of 370 to 900\,nm with $R\sim1800$ \citep{2012RAA....12.1197C,2012RAA....12..723Z,2012RAA....12.1197C}. Due to its distinctive structure, LAMOST can capture 4,000 spectra with a limited magnitude of {\it r = 18} in a single exposure \citep{2012RAA....12..723Z}. 
Over 11 million low-resolution spectra have been obtained, enabling the identification of many rare stars, such as Mira variables \citep{2017ApJS..232...16Y}, OB-type stars \citep{2019ApJS..241...32L}, and others, as reviewed by \citet{2016ApJS..226....1J} and \citet{2018ApJS..234...31L}.

\begin{figure*}
    \epsscale{1.0}
	\plotone{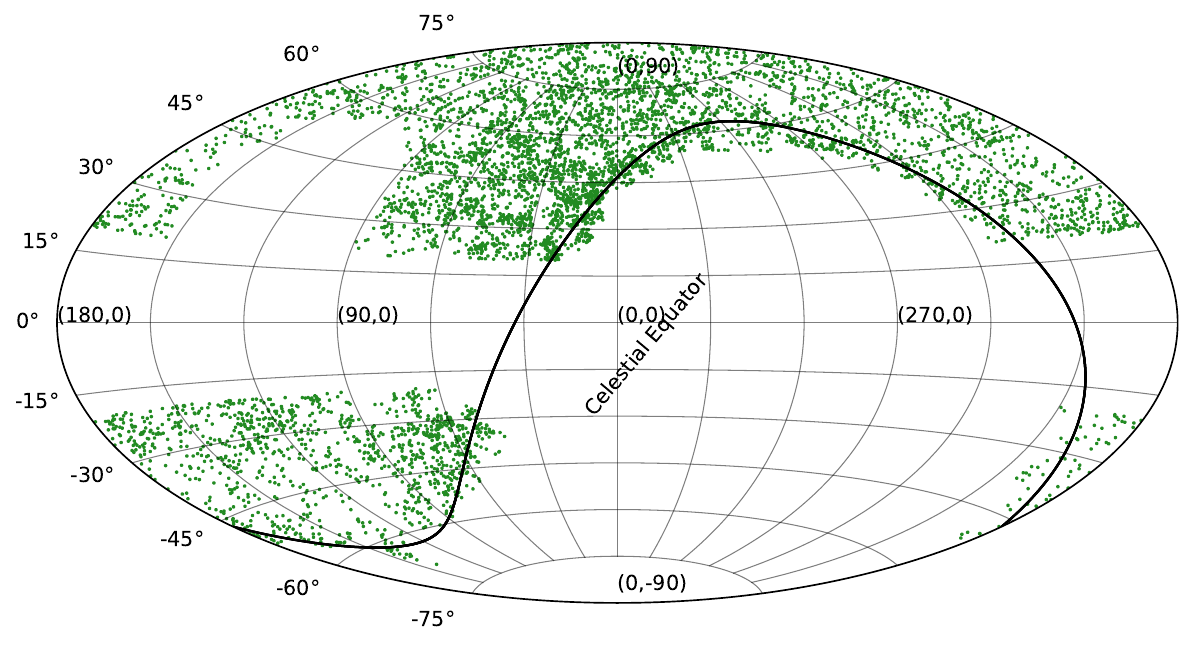}
	
	\caption{ The spatial distribution of BHB stars  identified from} LAMOST-LRS observations by \citet{2024ApJS..270...11J}   \label{fig:5355sp}
\end{figure*}

\citet{2024ApJS..270...11J} identified 5,436 BHB stars by measuring the equivalent widths of multiple absorption line profiles and spectral line features based on the LAMOST low-resolution spectra. Among them, 81 BHB stars do not have the color index of {\it Gaia}. Therefore, in this study, we adopt 5,355 BHB stars as our final sample. The spatial distribution of our sample stars is presented in Figure~\ref{fig:5355sp}. 

%

\section{Method}

\subsection{SLAM}

SLAM, a predictive stellar spectral model designed by \citet{2020ApJS..246....9Z,2020RAA....20...51Z}, has shown exceptional performance in spectral analysis. The model relies on the Support Vector Regression (SVR) algorithm \citep{smola2004tutorial,short2015improving} used in spectral data analysis. SLAM has three hyperparameters, the penalty level (C), tube radius ($\epsilon$), and the width of the radial basis function ($\gamma$), respectively. We have used the default hyperparameters as given by \citet{2020ApJS..246....9Z} ( c:[0.1,1,10];  $\epsilon$:0.05; $\gamma$:[0.1,1]).

Three main steps are needed when applying SLAM to the observed spectra data. Let's briefly summarize them:

1. Preprocessing: The first step is to preprocess the data. It involves normalizing the spectrum and standardizing each pixel of all normalized spectral fluxes and stellar parameters of the training data.

2. Training: After preprocessing, we use stellar parameters as independent variables and flux as dependent variables at each wavelength pixel to train the SVR model.

3. Prediction: Finally, we use the SVR model to predict stellar labels for observed spectra.

\begin{figure}
    \epsscale{1.2}
	\plotone{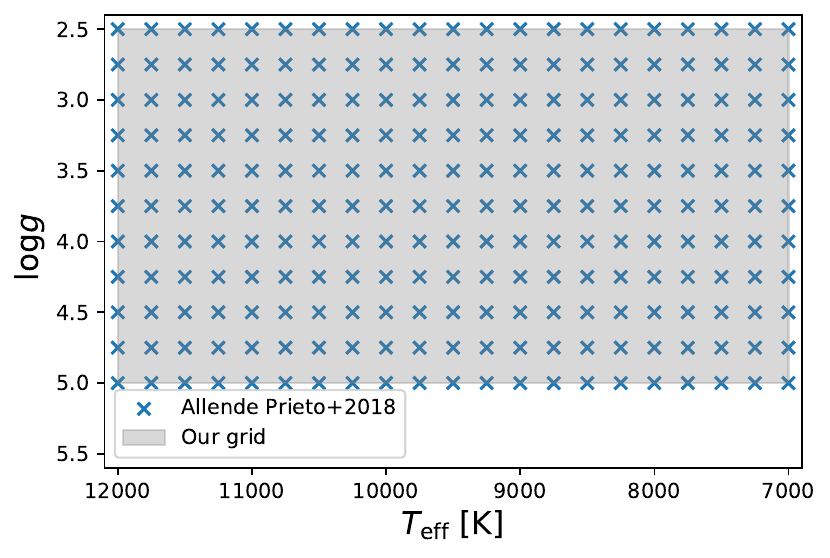}
	\caption{ The crosses in the ($T_{\rm eff}$~log~$g$) plane represents the model grids for A-type stars, while the grey-shaded area represents the linearly interpolated model grids that are employed for the training set. 
 \label{fig:ll}}
\end{figure}

\subsection{The Training Set}

Before performing SLAM on the spectra of BHB stars from LAMOST, it is essential to generate a training set, we conduct a thorough literature search to collect data on BHB stars with established stellar labels. However, the available samples do not cover the required parameter space for stellar labels. Based on our requirements, we adopt a set of theoretical spectra from \citet{2018A&A...618A..25A}, which provides a spectral grid with medium and high resolutions, covering the spectral type from B to early M. We use the 3200 A-type theoretical spectra from available grids of effective temperatures ranging from 7000 to 12,000\,K with a step of 250\,K, and surface gravities within 2.50 to 5.00 with a step of 0.25~dex. The metallicity range of the theoretical spectra is from $-$3.0 to 1.0 with a step of 0.2~dex. The wavelengths cover a range from 350 to 1000\,nm, and the sampling is not uniform. Figure~\ref{fig:ll} displays the grids for the theoretical spectra (crosses) in the ($T_{\rm eff}$, $\log{g}$) plane. 

To get the training set, we download the theoretical spectra from \citet{2018A&A...618A..25A}  and then generate additional theoretical spectra using a linear interpolation approach to increase the sample size. In principle, expanding the training sample by adding more spectral data could reduce the uncertainty in the stellar labels predicted from SLAM, but this takes quite a bit of computational time. Finally, we randomly select 5000 theoretical spectra for our training set according to \citet{2021ApJS..257...54G}. The enlarged grids are shown in the grey-shaded area in Figure~\ref{fig:ll}. To match the observed spectra obtained from LAMOST, we degrade the resolution of the theoretical spectra to $R\sim1800$  by Gaussian smoothing. We also resample the wavelength grids of the theoretical spectra to match the coverage of the observations, covering a wavelength range of 390-580\,nm with a step of 0.1\,nm. Meanwhile, we remove the defective pixels in the spectra.  Because the LAMOST low-resolution spectrum has a splicing of red and blue arms around 580\,nm and most of the spectral features are at the blue arm, so to save calculation time, the 580\,nm edge is selected as the cutoff wavelength for training.

\begin{figure*}
	\epsscale{1.2}
	\plotone{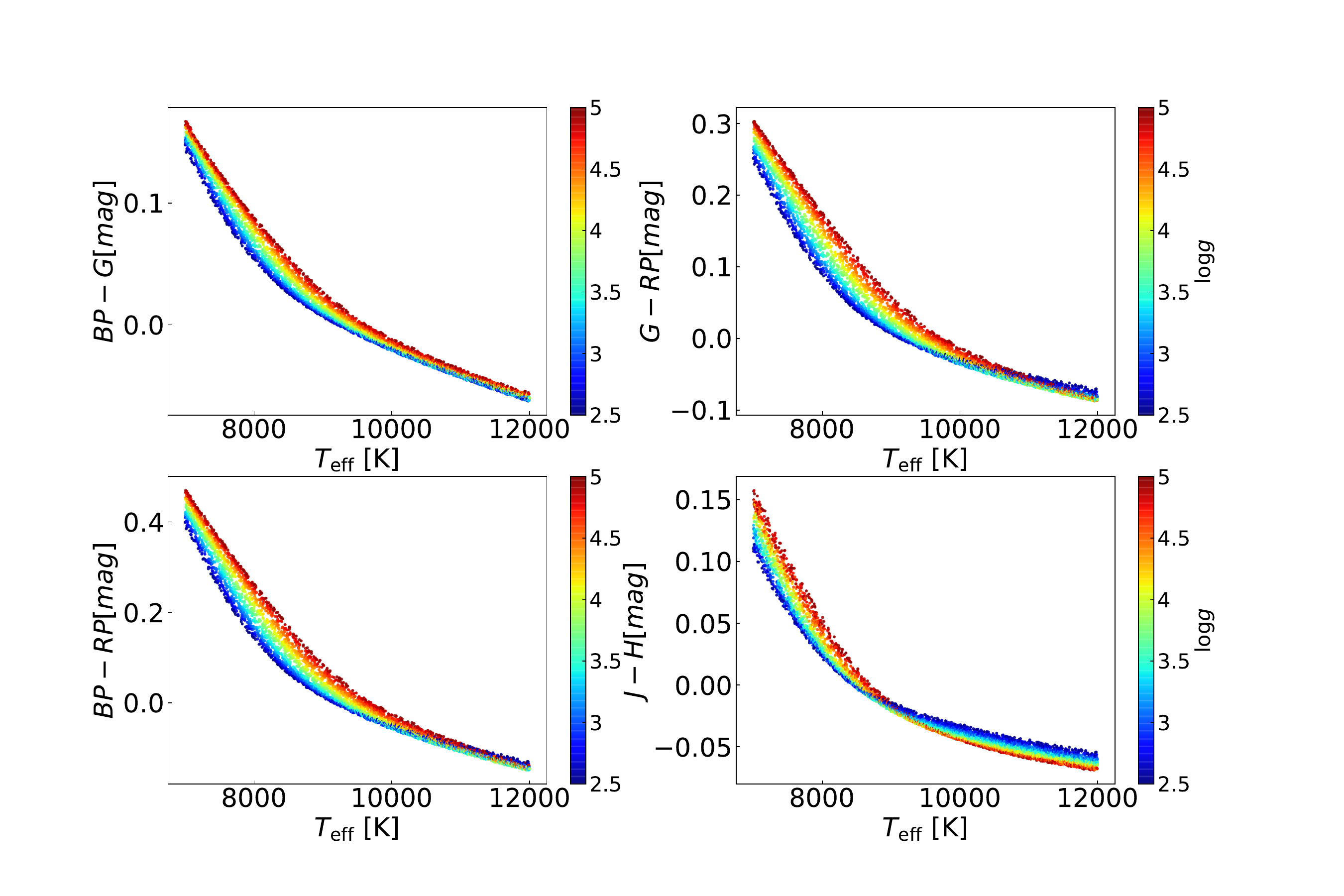}   
	\caption{The distributions of   $T_{\rm eff}$ vs. theoretical color indexes ($(BP-G), (G-RP), (BP-RP), (J-H)$) plane. The color bar on the right represents $\log{g}$. \label{fig:5000-t-color}}
\end{figure*}

It is important to point out that while the Balmer lines are crucial temperature indicators for A-type stars,  they do not change monotonously, reaching a maximum of about A2. Relying solely on the Balmer lines could lead to biased temperature estimations. The color index can effectively serve as a temperature metric. Previous studies have used color index to represent the temperatures \citep{2000ApJ...540..825Y,2004AJ....127..899S,2011MNRAS.416.2903D}. Figure~\ref{fig:5000-t-color} shows the anti-correlation between the temperature and four color indexes ($(BP-G), (G-RP), (BP-RP), (J-H)$). The color bar on the right represents $\log{g}$. The temperature monotonically increases with the decreasing of four color indexes, while the impact on $\log{g}$ is relatively small. Therefore, to more accurately assess the stellar atmospheric parameters, we adopt these four color indexes as four pixels to be added to the training sample along with the flux. 

\begin{figure*}
	\epsscale{1.2}       
	\plotone{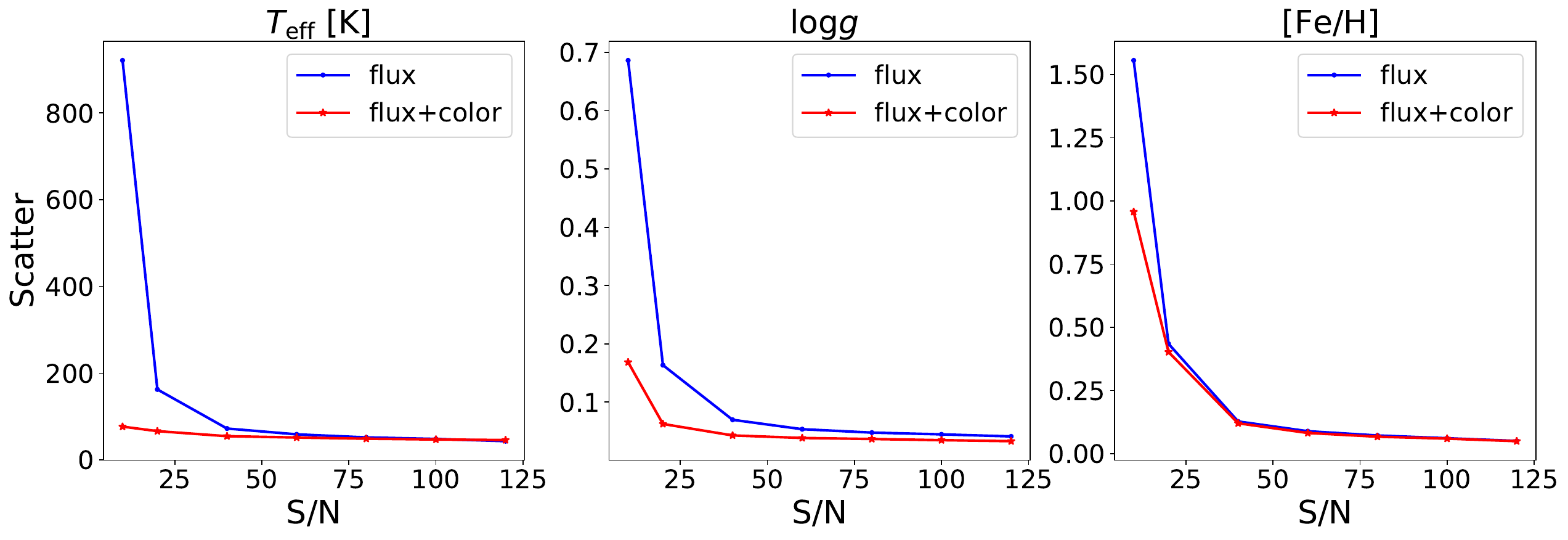}
	\caption{The distribution of Scatter values for predicted atmospheric parameter
 of $T_{\rm eff}$ $\log{g}$, and [Fe/H] as a function of S/N values. Blue represents training flux only, and red represents training flux and color index.\label{fig:scatter}}
\end{figure*}

\subsection{Self-consistency Check}

Before using SLAM to predict the stellar labels of the BHB stars identified in the LAMOST database, it is important to confirm that the module is reliable when applied to the training dataset. This validation process uses cross-validation (CV), known as k-fold cross-validation, which involves a consistency check. This widely accepted method evaluates the efficiency of machine-learning models more objectively and realistically. 

We follow the five-fold CV technique proposed by \citet{ojala2009permutation}. This method divides the set of spectra into 5 groups, and 4 of them are randomly selected as the training dataset. In contrast, the remaining spectra in the 5-set are used as the validation set. SLAM is then used to predict the stellar labels of the spectra in the validation set.

The scatter (standard deviation) is adopted to evaluate CV performance. It is defined as: 
\begin{equation}\label{eq:sersic1}
\text {Scatter }=\frac{1}{n} \sqrt{\sum_{i=1}^{n}\left(\theta_{i, \mathrm{SLAM}}-\theta_{i}\right)^{2}}
\end{equation}  
where, $\theta_{i, \mathrm{SLAM}} $ represents the estimated stellar label ($T_\mathrm{eff}$, $\log{g}$, [Fe/H]) of the ${i}th$ star by SLAM, and $\theta_i$  represents the true stellar label of the ${i}th$ star. The scatter values are 45\,K, 0.03~dex, and 0.04~dex for $T_\mathrm{eff}$, $\log{g}$, and [Fe/H], respectively. 

The signal-to-noise ratio (S/N) of input spectra affects the scatter value. To evaluate SLAM's performance on the spectra of different S/N, we apply a Gaussian function (mean = 0, $\sigma$ = flux/(S/N) ) to add a range of S/N values (from 10 to 120) to each pixel of the theoretical spectra. 
In Figure~\ref{fig:scatter}, we show the distributions of scatter values for each predicted stellar label as a function of S/N. It can be seen that when S/N$<$ 40, the scatters of the stellar label ($T_\mathrm{eff}$, $\log{g}$, [Fe/H]) are considerably reduced when the color index is included in the training compared to those of training with fluxes only. This reduction is particularly noticeable for the effective temperature. The scatter of $T_\mathrm{eff}$ for fluxes alone reaches 920\,K when S/N = 10, while it is only 76\,K when both the fluxes and color index are considered.  We use the reciprocal of the square of the error as the weight. This means that when the S/N of spectra is low, the color index has a larger weight and dominates the temperature estimation, ensuring a reliable temperature estimate even under low S/N conditions. $T_\mathrm{eff}$ and $\log{g}$ are coupled, which means that a reliable $\log{g}$ can also be predicted. However, the difference is not so significant when the S/N of a spectrum is higher than 40. It can also be seen that the improvement in metallicities is not obvious. 
Among our 5355 BHB stars, the spectra of 2229 objects have S/N  below 40, suggesting that adding the color index is necessary.

\section{ Results and Validation}

We obtain a set of spectra for 5355 BHB stars from the LAMOST DR5 database, and they are normalized using the Python package \textit{laspec}\footnote{ \href{https://github.com/hypergravity/laspec/}{https://github.com/hypergravity/laspec/}}. Then, we need to shift the LAMOST spectra to the rest frame using the radial velocity provided by the LAMOST DR5 database.  We crossmatch our sample with the {\it Gaia} EDR3 catalog \citep{2021A&A...650C...3G} and obtain their {\it BP}, {\it G}, and {\it RP} magnitudes. Meanwhile, we crossmatch with 2MASS \citep{2006AJ....131.1163S},  obtaining 4783  common stars of {\it J}, and {\it H} 
magnitudes.  The extinction law from \citet{2019ApJ...877..116W} is adopted for the four color indexes $((BP-G), (G-RP), (BP-RP), (J-H))$, and the reddening values from the 3D reddening map of \citet{2019ApJ...887...93G} is taken. 

\begin{figure*}
	\epsscale{1}
	\plotone{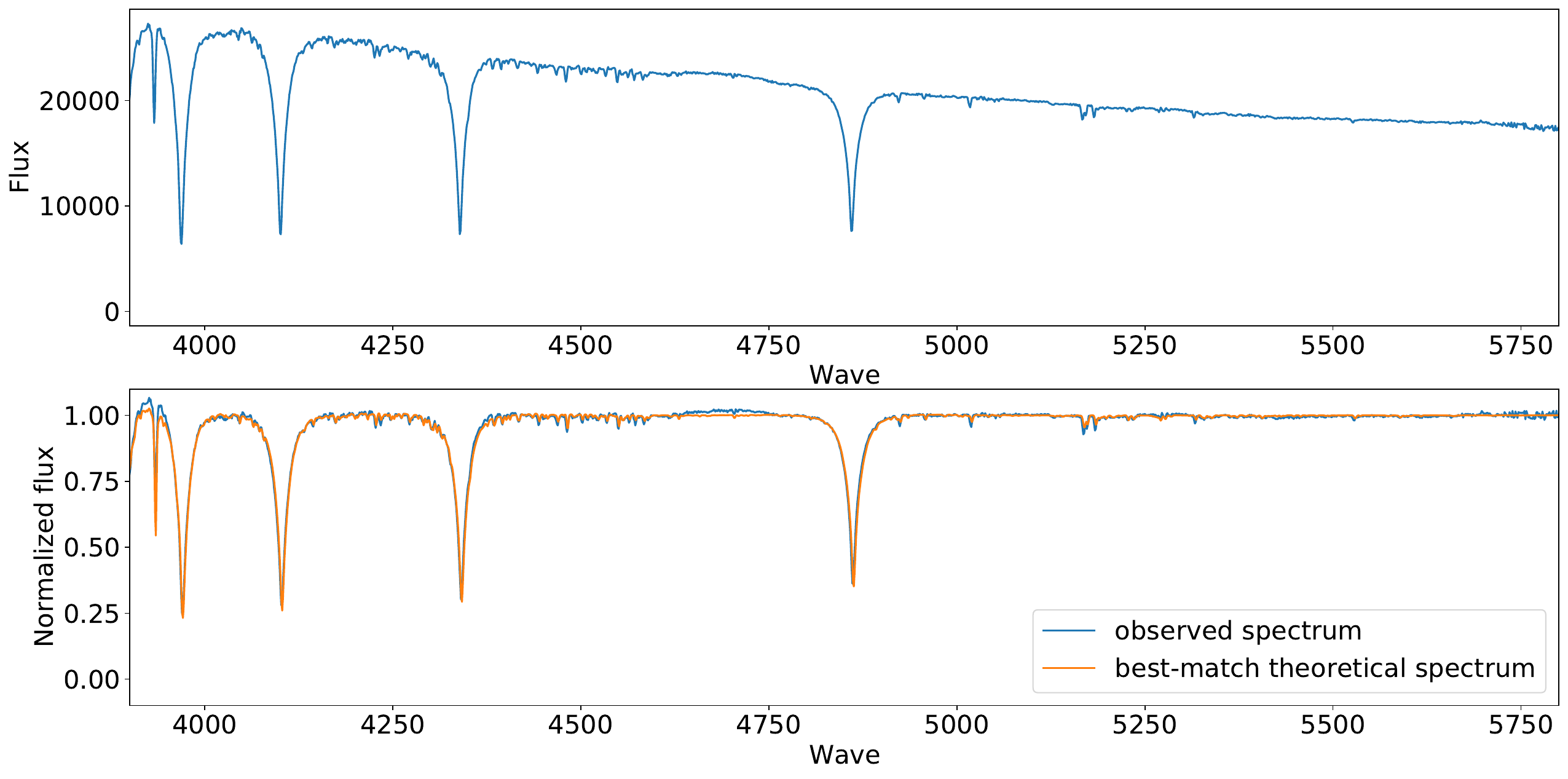}
	\caption{The upper figure shows the LAMOST low-resolution spectrum, the bottom shows the normalized and radial velocity corrected spectrum (blue line) and the theoretical spectrum (orange line) obtained based on the atmospheric parameters ($T_\mathrm{eff}$ = 7783\,K, $\log{g}$ = 3.23, and [Fe/H] = $-$1.63) from SLAM.  \label{fig:lamost_ll}}
\end{figure*}

In Figure~\ref{fig:lamost_ll}, the upper panel displays the LAMOST spectrum with OBSID\,392607019, while the bottom panel shows the spectra that have been normalized and corrected for the radial velocity (blue lines). According to its predicted stellar labels by SLAM, the stellar atmospheric parameters are estimated as $T_\mathrm{eff}$ = 7783\,K, $\log{g}$ = 3.23, and [Fe/H] = $-$1.63. We also plot the theoretical spectrum, shown as the orange line in Figure~\ref{fig:lamost_ll} for an example. 

\subsection{The Atmospheric Parameters of BHB stars }

We can predict the stellar labels of the sample  BHB stars using SLAM, and Table~\ref{Table 1} shows the estimated stellar parameters, including $T_\mathrm{eff}$, $\log{g}$, and [Fe/H]. The observed ID, coordinates and the S/N of these spectra are also presented in Table~\ref{Table 1}.  

\begin{figure*}	             
    \epsscale{1.3}
	\plotone{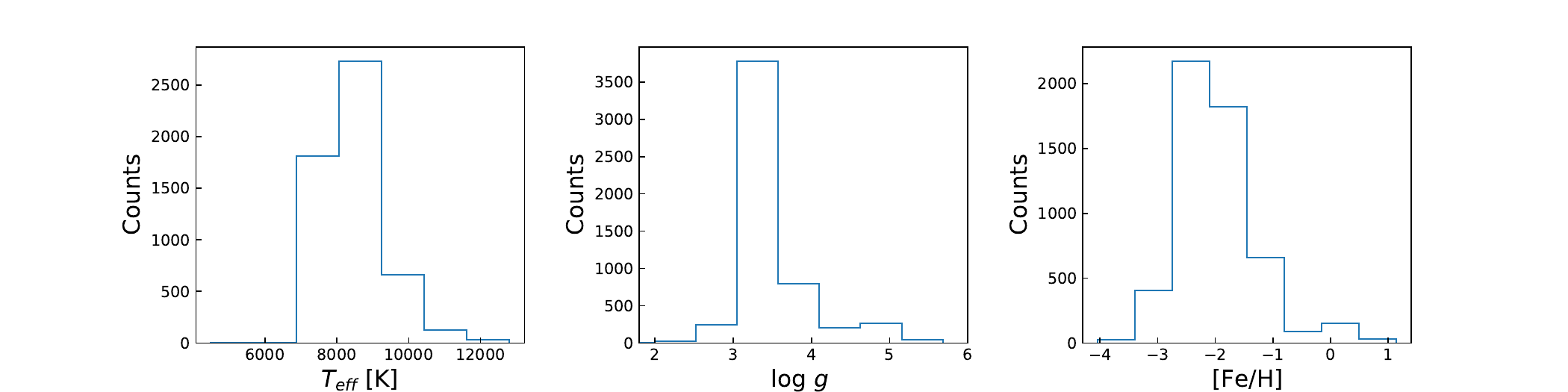}
	\caption{The distribution of BHB stars in $T_\mathrm{eff}$  (left panel), $\log{g}$  (middle panel)} and [Fe/H] (right panel). 
	\label{paramas_hist}
\end{figure*}

\begin{figure*}
	\epsscale{1.1}
	\plotone{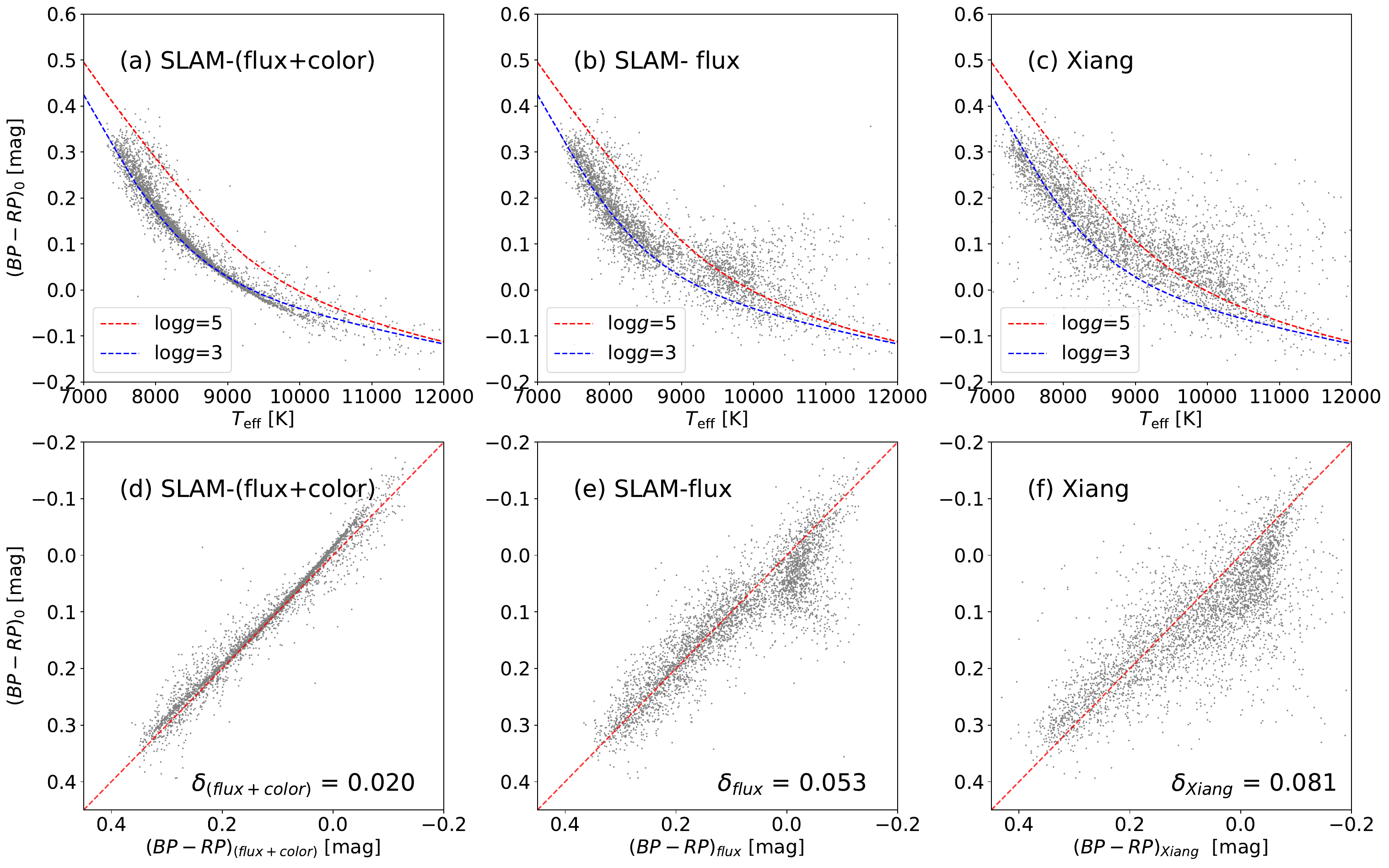}
	\caption{The distribution of BHB stars in the $T_\mathrm{eff}$ vs. $(BP-RP)_0$ plane from panel (a) to panel (c). $T_\mathrm{eff}$ is estimated by training flux and color index(panel (a)), by training flux(panel (b)), by Xiang(panel (c)). The dashed line represents the theoretical results predicted by the PARSEC models with [Fe/H]$= -2.0$, $\log{g}$ = 3 (blue dashed line) and 5 (red dashed line), respectively. Panel(d)-(f) compares the theoretical $(BP-RP)_0$ calculated from the predicted parameters and the observed color indices($(BP-RP)_0$) after extinction correction.
 \label{fig:t-bprp0}}
\end{figure*}

Figure~\ref{paramas_hist} presents the distribution of atmospheric parameters for the BHB stars. The $T_\mathrm{eff}$ ranges from 7000\,K to 12,000\,K, the range of $\log{g}$ is about 2.5 to 5.5, and the metallicity is concentrated at $-$2.0. In Figure~\ref{fig:t-bprp0}, we plot the predicted $T_\mathrm{eff}$ by SLAM versus $(BP-RP)_0$.  It can be seen that $T_\mathrm{eff}$ is correlated with $(BP-RP)_0$ from (panel (a), but there is still a scatter, especially at $T_\mathrm{eff} < $ 9000\,K. The $T_\mathrm{eff}$ (7000\,K to 12,000\,K)-dependent trajectory on the synthetic color from the PARSEC models with [Fe/H]$= -$2.0 and $\log{g}$ $= $ 3 (blue), and 5 (red)\citep{2019A&A...632A.105C} are also shown. Since metallicity has little effect on $(BP-RP)_0$, we only show the effect of $\log{g}$ on $(BP-RP)_0$. For different $\log{g}$, $T_\mathrm{eff}$ is different when $(BP-RP)_0$ is the same, especially at low temperatures. This result indicates that the scatter is real. 

\begin{figure*}
	\epsscale{1.2}
	\plotone{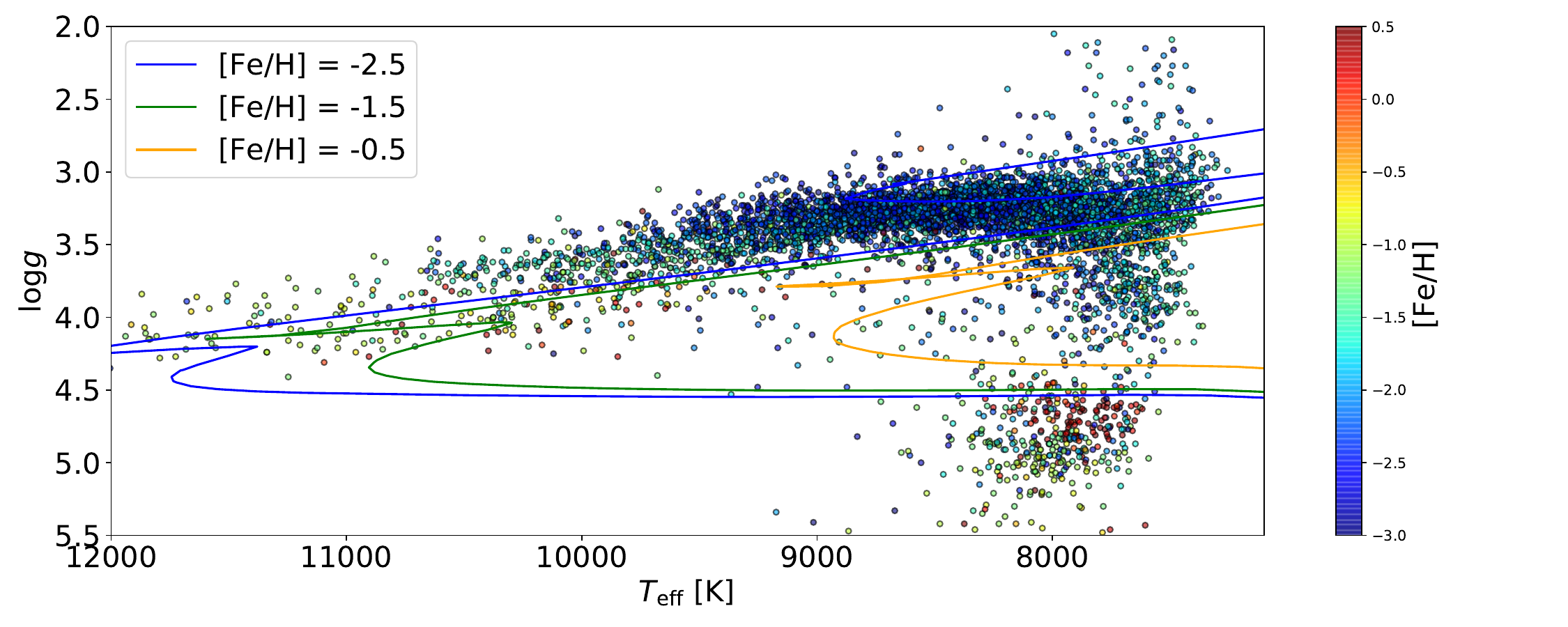}
	\caption{ The distribution of BHB stars in the $T_\mathrm{eff}$ vs. $\log{g}$ plane. The lines represent PARSEC, the theoretical isochrone tracks, with metallicity values of [Fe/H] = $ -$2.5,$-$1.5, and $-$0.5, respectively. The tracks with ages of 1 Gyr are included in the figure. \label{fig:t-logg}}
\end{figure*}

\begin{figure*}
	\epsscale{1.0}
	\plotone{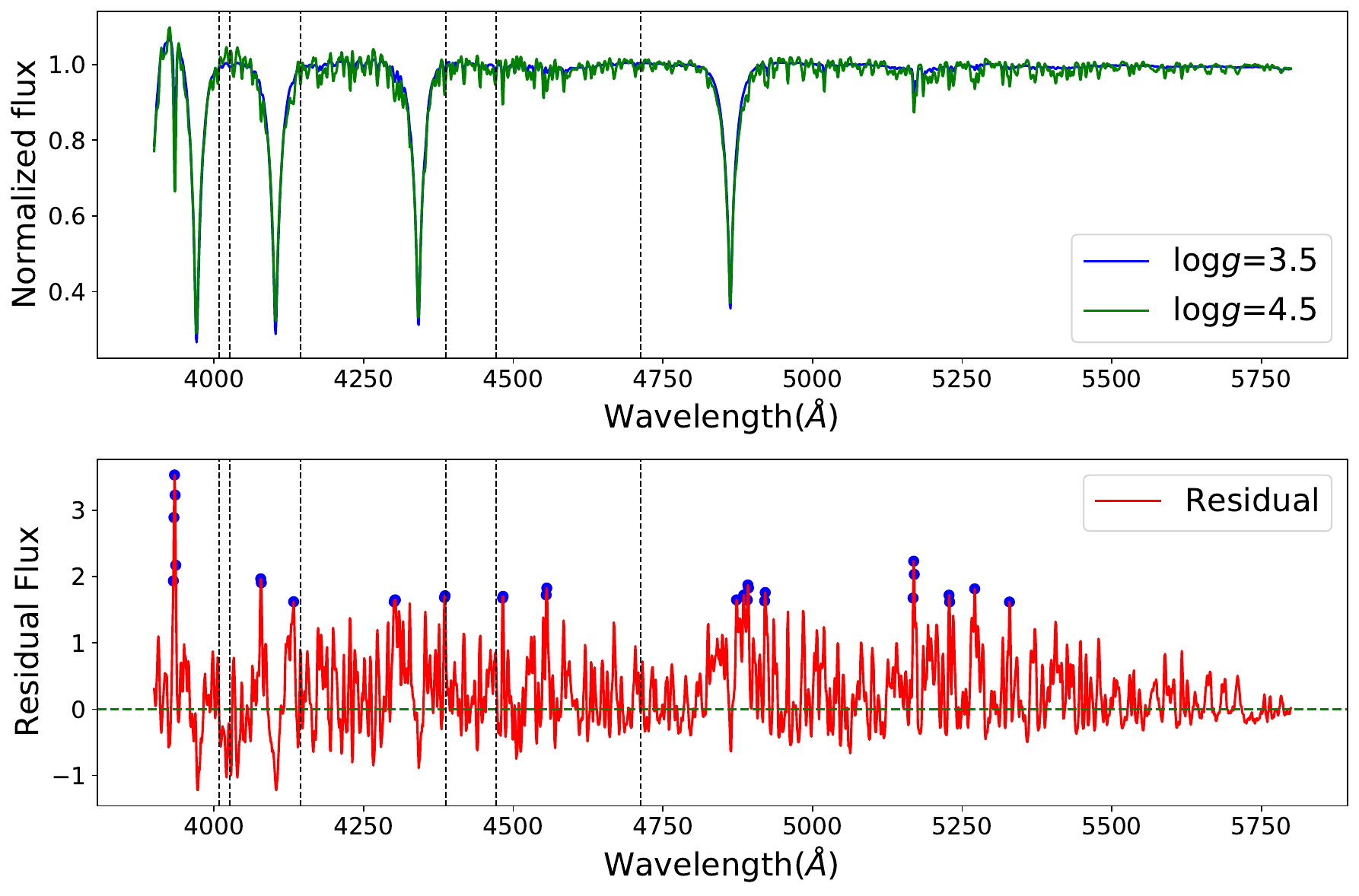}
	\caption{ The upper panels compare average BHB spectra with large and small $\log{g}$ around t=8000K. The bottom panels show the normalized residuals between
these two average spectra, the blue dots represent outliers exceeding three standard deviations. The black dashed lines represent the He absorption lines. \label{fig:compare_logg}}
\end{figure*}

Figure~\ref{fig:t-logg} displays the predicted $T_\mathrm{eff}$ as a function of $\log{g}$, with the color bar indicating metallicities. We derived the isochrones from the Padova and Trieste Stellar Evolutionary Code (PARSEC) \citep{2000MNRAS.315..543H,2012MNRAS.427..127B}, and overplotted them with the ages of 1 Gyr in the figure. Three sets of isochrones with values of [Fe/H] = $-$2.5 (blue lines), [Fe/H] =  $-$1.5 (green lines), and [Fe/H] =  $-$0.5 (orange lines) are shown for the age tracks. The theoretical isochrones fit well with the predicted stellar labels for the observed spectra given by the SLAM.  Approximately 350 BHB stars show $\log{g}$ $>$ 4.5, higher than most BHB stars with comparable temperatures ranging from 7000 to 8000\, K. Among these, around 150 BHB stars also display higher metallicities ([Fe/H] $>-$1.0), some even approaching 0.5. We have plotted the average observed spectra of stars with $\log{g}$ $>$ 4.5 (in green) and normal $\log{g}$ (in blue) around 8000 K and calculated the normalized residuals between these two average spectra(Figure~\ref{fig:compare_logg}). Outliers exceeding three standard deviations are marked in blue dots, and common He lines (4009Å, 4026Å, 4144Å, 4387Å, 4471Å, 4713Å) are indicated with black dashed lines. He I (4387Å) line among these outliers, suggesting a possible association with helium abundance issues. However, this line constitutes a small fraction of the outliers, and the majority of the spectral residuals are not attributed to helium abundance\citep{2009Ap&SS.320..261C}. there may be other potential parameters that could account for these phenomena, their effects are likely to be secondary.  Additionally, we cross-match with the variable star catalog \citep{2016AcA....66..405S} and identified 32 variable stars. At the same time, we also see that a small number of 320 stars in the sample with the predicted stellar labels lie outside the restricted parameter range of our training set, especially $\log{g}$ and metallicity. We visually examine the spectra of these stars and find that they are either spectra in poor quality in which the \ion{Ca}{2}\,K line is blended by noise or special spectra with emission lines. The temperature is in the range because we use a wide band color index $(BP-RP)$. When the spectrum is special, the color index can still be a good temperature indicator. However, $\log{g}$ and metallicity are unreliable. We also found that in the high-temperature range, there are about a hundred BHB stars with significantly increased metallicity, consistent with the literature suggesting that BHBs are affected by atmospheric diffusion phenomena when $T_\mathrm{eff}$ $>$11000\,K \citep{behr2003chemical}. However, the predicted metallicity does not reach the solar metallicity, indicating a certain degree of deviation in the predicted results for these stars. In the future, we will consider incorporating diffusion-stratified atmospheric models to more accurately predict the atmospheric parameters of BHBs within this temperature range.

\begin{figure}
	\epsscale{1}
	\plotone{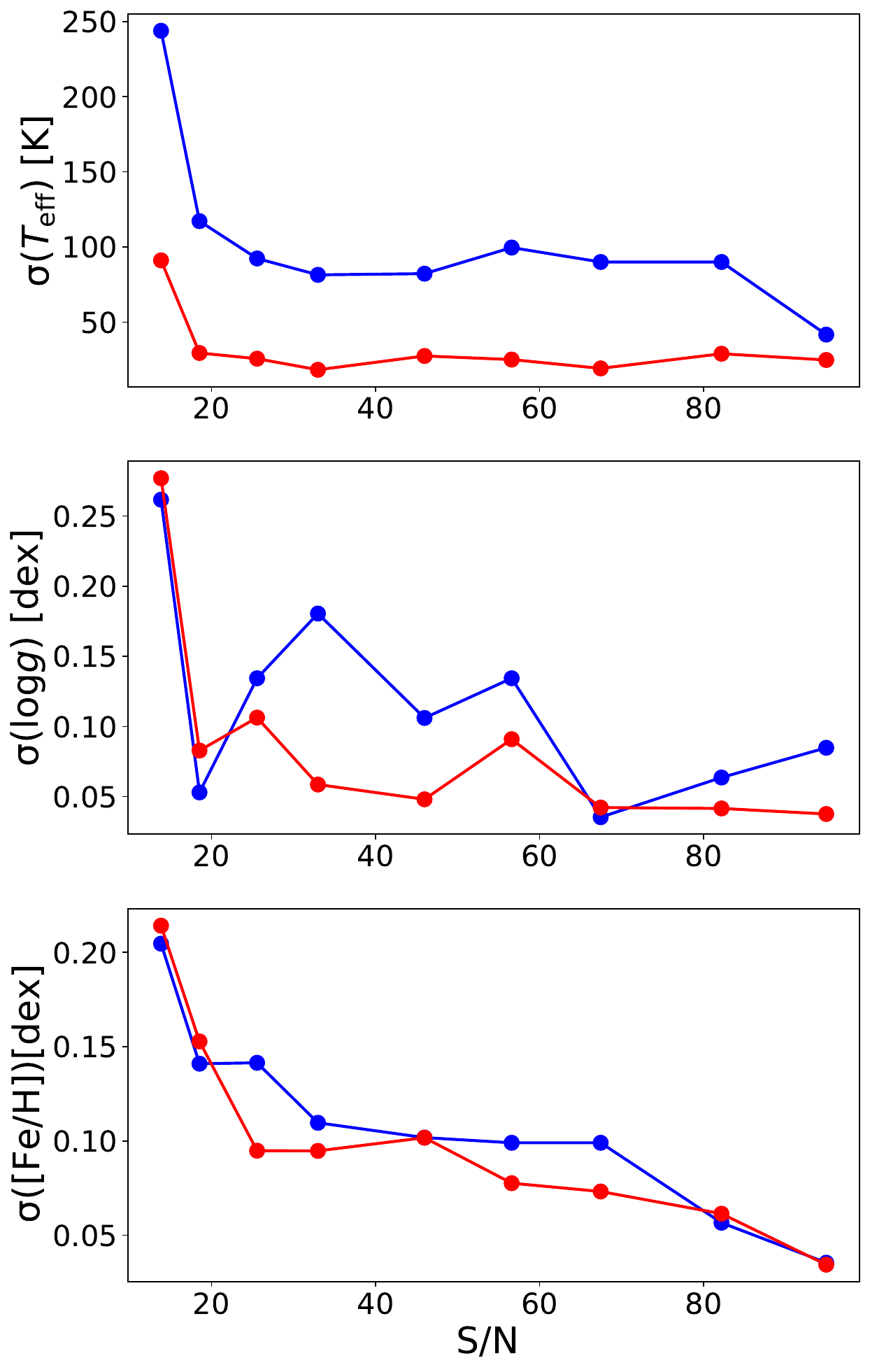}
	\caption{ The random errors of stellar parameters for duplicate as functions of S/N. The blue line represents training flux only, and the red line represents training flux and color index.   \label{fig:snr}}
\end{figure}


\begin{figure*}
	\epsscale{1.4}
	\plotone{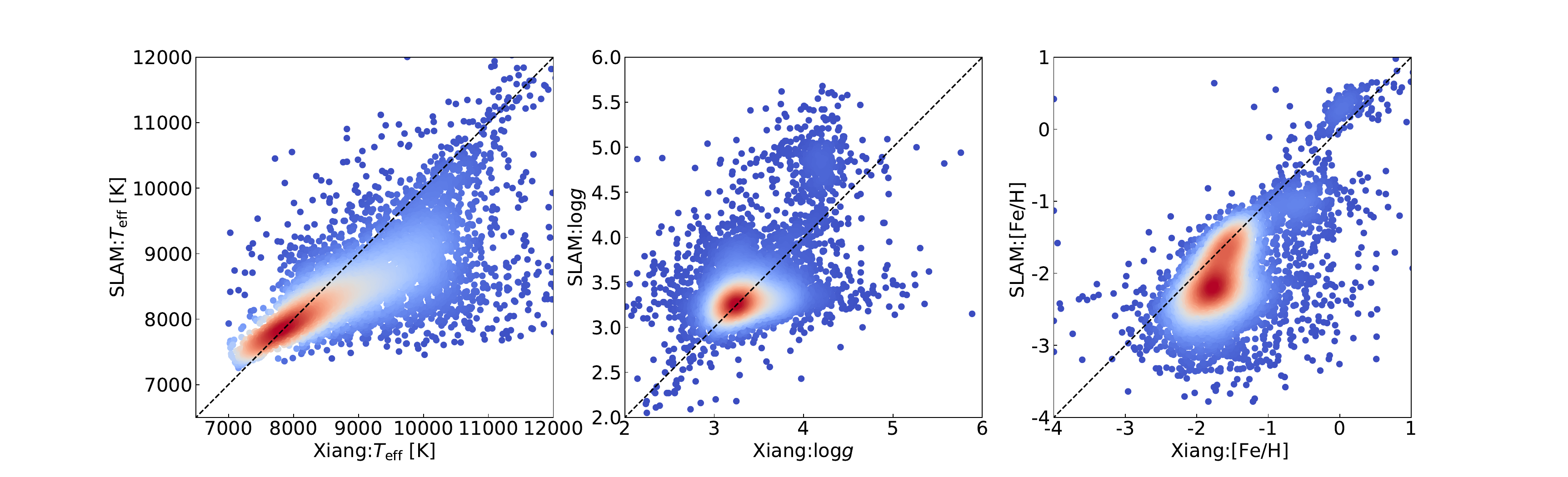}
	\caption{  Comparison of the stellar parameter between the SLAM estimates and Xiang's results for the BHB stars.   \label{fig:t-compare_xiang}}
\end{figure*}

\begin{table}
\footnotesize
\centering
    \caption{The Predicted Stellar Labels of 5355 LAMOST BHB Stars. }\label{Table 1}
    \begin{tabular}{ccccccccc}
    \hline
    Index &        Label (FITS) &        Format &     Units &     Description  \\
     
    \hline
1&	spid&	Integer&	...&	Spectrograph ID\\
2&	ra&	Double&	deg&	R.A.(J2000)\\
3&	dec&Double&	deg&	Decl.(J2000)\\
4&	$S/N_g$&	Float&	...&	S/N of g band\\
5&	$T_{\rm eff}$&	Float&	K&	Predicted by SLAM \\
6&	log\,$g$&	Float&	...&	Predicted  by SLAM\\
7&	[Fe/H] &	Float&	...&	Predicted by SLAM\\

    \hline& 
    \end{tabular}
    \tablecomments{Table 1 is published in machine-readable (FITS) format. A portion is shown here for guidance regarding its form and content.
(This table is available in its entirety in FITS format.)}
    
    \end{table}

\subsection{Multiple observations}

Numerous stars have had multiple visits during the LAMOST survey, providing valuable insights into the random errors of atmospheric parameters. There are 1116 objects in our sample with multiple observations. The star with the highest number of repeated observations has seven individual spectra, while the median number of observations per star is two.

 We only consider the spectra of repeated observations with a difference in S/N less than 20\%. In Figure~\ref{fig:snr}, the standard dispersion of stellar parameters for these objects of multiple visits are shown as a function of S/N.  The standard dispersion represents the errors of stellar parameter determination, which show a decreasing tendency as the S/N of spectra increases. The error of only the flux considered in training is also plotted (blue line). The error of the training set with the color index included is significantly smaller than that of the training set without the color index considered, but the difference in the error of metallicity is small.

The S/N influences the derived effective temperatures. When S/N is around 10, the precision of $T_{\rm eff}$ predicted by the training including both the flux and color index (red line) is approximately 100\,K, and it decreases to 30\,K when  S/N $>$ 20. However, the precision of $T_{\rm eff}$ estimated by only the flux considered in training (blue line) is about 250\,K, and it decreases to 100\,K when the S/N of spectra $>$ 20. When the S/N of spectra is higher than 20, the determined temperature is less affected. It is due to the color index does not change between different spectra of the same star. When the S/N of a spectrum is relatively low, the color index has a stronger impact. Surface gravities are also affected by the S/N of spectra, possibly due to the difficulty in deriving them for the noise-induced spectra (broadening of the Balmer lines). The determination of $\log{g}$ can achieve a precision better than 0.1 for the spectra of S/N greater than 20. However, for the low-S/N spectra, the precision on measurement of $\log{g}$ is poor, with a dispersion that can be as high as 0.25. A similar trend can be found for metallicities,  there is a notable correlation to S/N. The precision of [Fe/H] is approximately 0.17 when the spectra of S/N$\sim$10, but decreases with increasing S/N. Once the S/N exceeds 40, the precision can be lower than 0.1~dex. The error still has a decreasing trend when the S/N is even higher than 80.


\subsection{Comparison with training without color index}

 Figure~\ref{fig:t-bprp0} panel(b) displays the relationship between the estimated $T_\mathrm{eff}$ by training only using flux and $(BP-RP)_0$. For comparison, we also show the predicted colors using the PARSEC models with [Fe/H]$= -$2.0 and $\log{g}$ $=3$ (blue dashed line), 5 (red dashed line), respectively \citep{2019A&A...632A.105C}. It can be seen that for most of our sample stars with $T_\mathrm{eff}<8500$~K their temperatures and $(BP-RP)_0$ show basically consistency with the theoretical results of metal-poor stars at their BHB stage. However, large scatters of $(BP-RP)_0$ exist at a fixed $T_\mathrm{eff}$, especially in the region with $T_\mathrm{eff}>9000$~K (see Figure~\ref{fig:t-bprp0} panel(b)). The strength of Balmer lines for A-type stars reach their maximum around A2. If only the spectral flux of Balmer lines (the major absorption lines in the spectra of hot stars) is used in training, severe temperature degeneracy may occur. Because there are correlations between temperatures and color indexes, which can help to break temperature degeneracy. By comparison, we can see that the temperature obtained by adding color index in training has a better consistency to the theoretical $(BP-RP)_0$ with small scatters (see Figure~\ref{fig:t-bprp0} panel (a)). We have calculated the theoretical color indices $(BP-RP)$ based on the atmospheric parameters estimated with color index (Figure~\ref{fig:t-bprp0} panel(d)) and without color index (Figure~\ref{fig:t-bprp0} panel(e)). We compared these with the observed color indices by applying stringent constraints as shown in Figure 3: $(BP-RP)_0 < 0.4$, $\text{err}_{\text{BP-RP}} < 0.01$, and $E(B-V) < 0.1$. The results show that $\sigma_{(flux+color)} = 0.020$ and $\sigma_{flux} = 0.053$, indicating that SLAM training with color index provides more accurate temperature predictions.

\subsection{Comparison with previous work}

\citet{2022A&A...662A..66X} estimated the atmospheric parameters and elemental abundances of hot OBA stars using the HOTPAYNE method from the LAMOST-LRS. We cross-match their data with our sample stars and find 5297 common objects. In Figure~\ref{fig:t-compare_xiang}, we compare the star labels ($T_\mathrm{eff}$, $\log{g}$, [Fe/H]) of SLAM and with their work. Both results are consistent when the temperature is between 7000 - 9000\,K. However, when $T_\mathrm{eff} >$ 9000\,K, the temperature of some stars predicted by \citet{2022A&A...662A..66X} is higher than ours,
although they have large $(BP-RP)_0$ values (see Figure~\ref{fig:t-bprp0} panel (c).

 The large deviation in metallicity for both works may be due to the weak metal lines of the hot BHB stars, especially when the temperature is higher than 8500\,K. Therefore it is difficult to derive a reliable metallicity.

 Figure~\ref{fig:t-bprp0} panel (f) compares the theoretical $(BP-RP)_0$ calculated from the predicted parameters and the observed color indices,$(BP-RP)_0$, after extinction correction. We thus calculated the standard deviation $\sigma_{Xiang}$ = 0.081. And ours is only 0.020, indicating our results are more reliable. 

\begin{figure*}
	\epsscale{1}
	\plotone{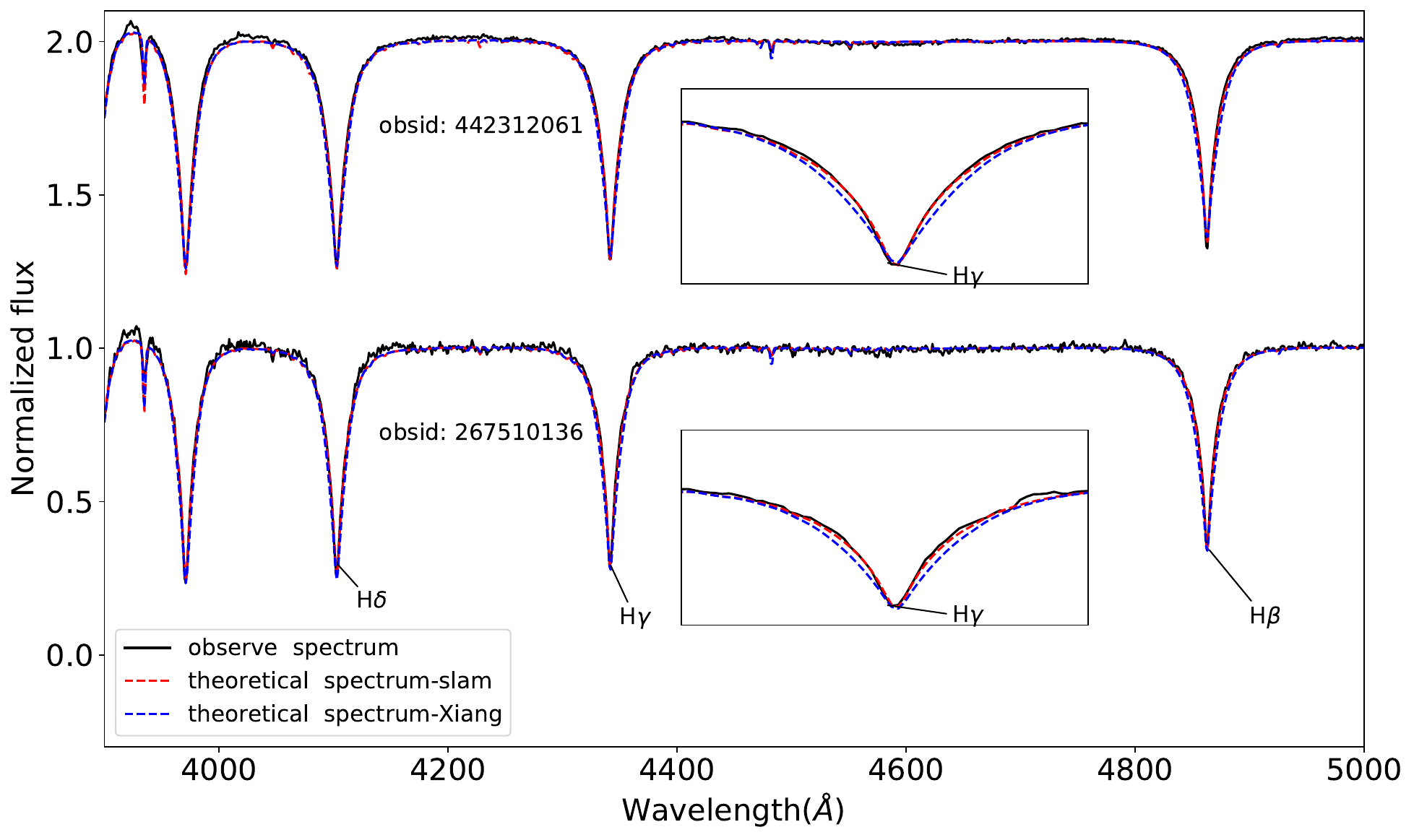}
	\caption{The normalized spectrum of two stars. The black lines show the observed spectrum of the star.
The red dashed line shows the theoretical spectra.  \label{fig:compare_spectra}}
\end{figure*}

We select two common stars showing large differences in atmospheric parameters and compare them in Table~ \ref{Table2}. Figure~\ref{fig:compare_spectra} displays the observed high S/N spectra (black) and theoretical spectra for these two stars (red represents the atmospheric parameters predicted by SLAM, and blue is by \citet{2022A&A...662A..66X}). It can be seen that the results predicted by SLAM fit the observed spectra better especially for the Balmer line wings. Meanwhile, our temperature well matches the color index $(BP-RP)$ with the color index included in the training,  which indicates that our temperature is better estimated.

\begin{table*}
\footnotesize
\centering
    \caption{ }\label{Table2}
    \begin{tabular}{ccccccccc}
    \hline
    obsid &  $S/N_g$  &    $(BP-RP)_0$ &       $T_\mathrm{eff}$ &    $\log{g}$ &[Fe/H] &       $T_\mathrm{eff}$ &    $\log{g}$ &[Fe/H]\\
    &  & &(this work) &(this work) &(this work)  &(Xiang) &(Xiang)&(Xiang) \\
    \hline
    
    267510136&86.37   &0.22&7942.01\,$\rm K$   &3.29&-2.44&9334.3\,$\rm K$&3.79&-1.73\\
   442312061&306.31   &0.17&8100.41\,$\rm K$   &3.59&-2.3&10400.00\,$\rm K$&4.35&-1.64\\
       
    \hline 
    \end{tabular}
\end{table*}

\subsection{Comparison to High-resolution Spectra}

\begin{figure*}
	\epsscale{1.2}
	\plotone{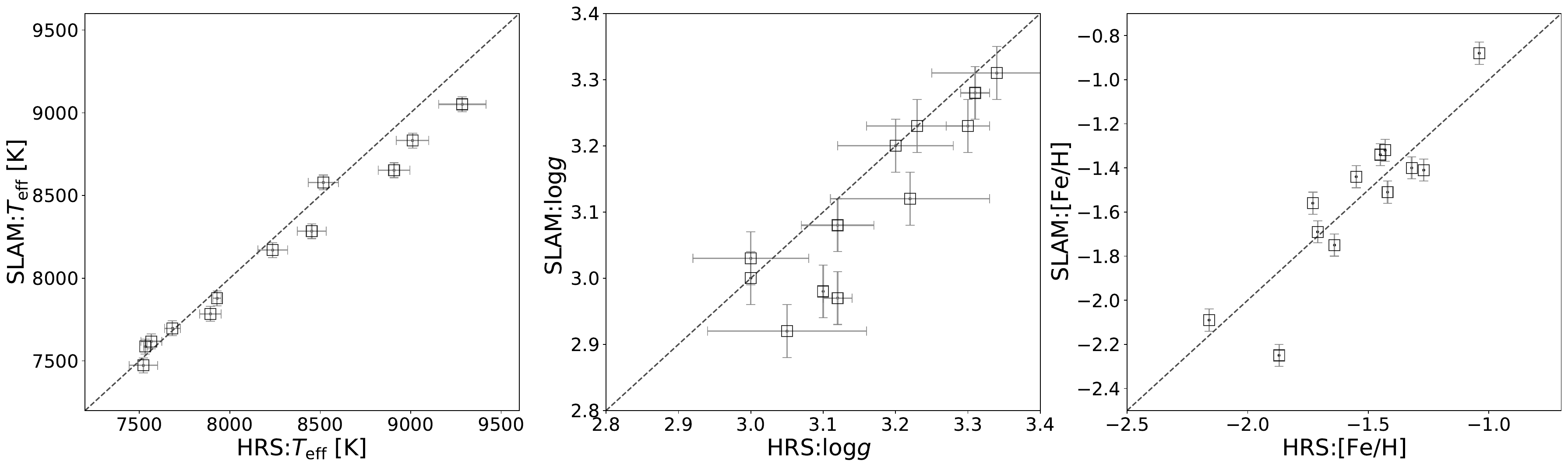}
	\caption{ Comparisons of predicted stellar labels ($T_\mathrm{eff}$, $\log{g}$, [Fe/H]) of 12 BHB stars obtained from the SLAM to the prelabeled values as given by \citet{1999AJ....117.2308W}.  \label{fig:eso12}}
\end{figure*}

We check our method using the high-resolution spectroscopy (HRS) samples from published data. We compile 12 BHB stars from \citet{2000A&A...364..102K} and predict their labels using SLAM. The LAMOST survey have not observed these stars, however, the high-resolution optical spectra are available in the scientific archives of the European Southern Observatory (ESO). For consistency, we reduce the resolution of spectra to the LAMOST one $R\sim1800$ (HRS-LRS). Figure~\ref{fig:eso12} compares the labels ($T_\mathrm{eff}$, log\,$g$, and [Fe/H]) estimated by SLAM with their reference values, and it can be seen that our results are in good agreement with the literature values within the uncertainty. The standard deviation between the predicted labels and the published values in the HRS are $\sigma$($T_\mathrm{eff}$) = 76\,K, $\sigma$(log\,$g$) = 0.04~dex, and $\sigma$([Fe/H]) = 0.09~dex, respectively. The sample size of the high-resolution BHB stars in our study is indeed limited. This may introduce a degree of randomness in our results and potentially underestimate the errors.  


\section{Discussion}

The stellar labels of stars are contained within their spectra, and stellar parameters are usually interrelated. For instance, hydrogen lines are commonly used to estimate effective temperature, and these lines are susceptible to surface gravity for A-type stars. The traditional method involves comparing observed and model spectra to determine stellar labels through minimization techniques, whereas SLAM simultaneously predicts the stellar labels of the input spectra.

\begin{figure*}
	\epsscale{1.0}
	\plotone{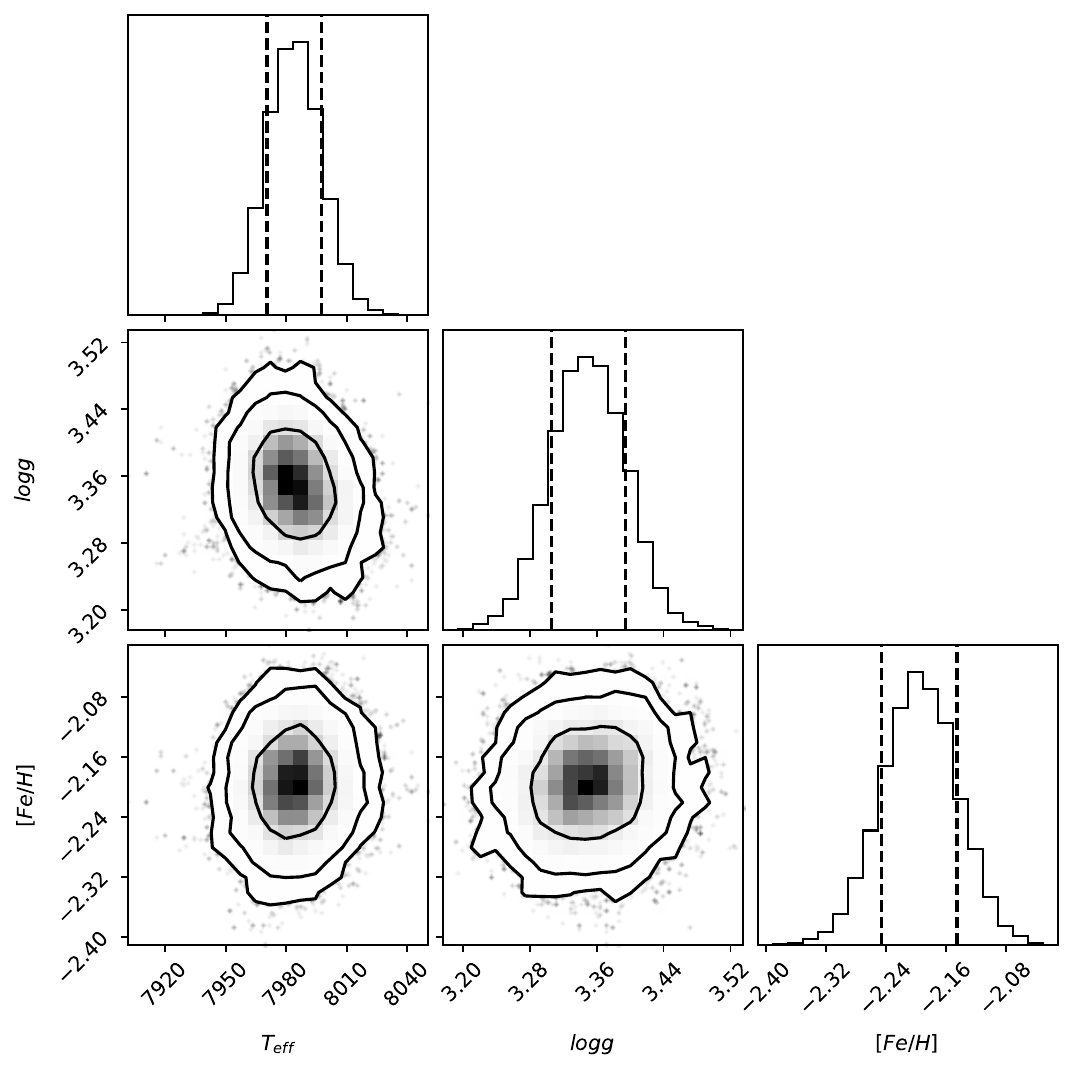}
	\caption{ Posterior distributions of predicted $T_\mathrm{eff}$ , $\log{g}$ and [Fe/H]. The contours from the inside out enclose the 68th, 95th, and 99th percentiles of the total
probability in the three off-diagonal panels. The two vertical dashed lines in the three diagonal panels represent the lower and upper errors of the distribution at the 16th
and 84th percentiles, respectively.  \label{fig:mcmc}}
\end{figure*}

To verify the potential degeneracy among the predicted stellar labels obtained from SLAM, we employed Markov Chain Monte Carlo (MCMC) simulations to examine the posterior distribution of the stellar labels. We present the results for the spectrum with a typical signal-to-noise ratio of 70.7 and observation ID OBSID 150705245 in the figure~\ref{fig:mcmc}. In the three off-diagonal, the contours delineating the 68\%, 95\%, and 99\% confidence regions are depicted from innermost to outermost as solid black lines. Within the three diagonal subplots, the two vertical dashed lines correspond to the lower and upper error bounds of the 16th and 84th percentile distribution for a given stellar label. The distribution indicates that no correlation is found between the three stellar labels
($T_\mathrm{eff}$ , $\log{g}$ and [Fe/H]). We consider that the predicted results are acceptable.

\section{Conclusions}

In this paper, we take 5000 random A-type theoretical spectra as a training data set and use the SLAM to predict the stellar atmospheric parameters of 5355 BHB stars from LAMOST DR5. Four color indexes ($(BP-G), (G-RP), (BP-RP), (J-H)$) are added to the training set in addition to spectral flux for getting with those of training without the color indexes, the temperature obtained by adding color index in training correlates better with the theoretical $(BP-RP)_0$. The predicted $T_\mathrm{eff}$ and $\log\,g$ can be improved when the color index is considered, especially for the spectrum with low S/N. 
We also predict the stellar atmospheric parameters of 12 high-resolution spectra of BHB stars to estimate realistic errors in the stellar labels predicted by SLAM. The realistic errors are $\sigma$($T_\mathrm{eff}$) = 76\,K, $\sigma$(log\,$g$) = 0.04~dex, and $\sigma$([Fe/H]) = 0.09~dex, respectively.  

It is demonstrated that the applicability of deriving stellar labels by adding the color index to a large sample of BHB stars using SLAM. The technique is expected to predict reliable stellar labels for A-type stars.

\begin{acknowledgments}
This study is supported by the National Natural Science Foundation of China under grant No. 12173013, 12090044, and 12203068; the project of Hebei Provincial Department of Science and Technology under grant number 226Z7604G, and the Hebei NSF (No. A2021205006). We acknowledge the China Manned Space Project. The Guoshoujing Telescope (the Large Sky Area Multi-Object Fiber Spectroscopic Telescope, LAMOST) is a National Major Scientific Project built by the Chinese Academy of Sciences. LAMOST is operated and managed by the National Astronomical Observatories, Chinese Academy of Sciences. This work has made use of data from the  European Space Agency (ESA) mission Gaia (\url{https://www.cosmos.esa.int/gaia}), processed by the Gaia Data Processing and Analysis Consortium (DPAC, \url{https://www.cosmos.esa.int/web/gaia/dpac/consortium}). Funding for the DPAC has been provided by national institutions, in particular, the institutions participating in the Gaia Multilateral Agreement. This study used the 2mass database to obtain J- and H-band magnitudes (\url{https://irsa.ipac.caltech.edu/Missions/2mass.html}).
\end{acknowledgments}

%

\vspace{5mm}
\facilities{LAMOST}





\bibliography{ref}{}

\end{document}